\documentclass[english]{JHEP3}
\usepackage{amsmath}
\usepackage{amssymb}

\makeatletter

\usepackage{epsfig}
\usepackage{multicol}
\usepackage{multirow}
\usepackage{subfigure}
\usepackage{latexsym}
\usepackage{cite}
\usepackage{here}
\usepackage{bbm}

\newcommand{\fverb}{\setbox\fverbbox=\hbox\bgroup\verb}
\newcommand{\fverbdo}{\egroup\medskip\noindent%
\fbox{\unhbox\fverbbox}\ }
\newcommand{\fverbit}{\egroup\item[\fbox{\unhbox\fverbbox}]}
\newbox\fverbbox

\allowdisplaybreaks

\title{Power-law expansion of the Universe from the
bosonic Lorentzian type IIB matrix model}

\author{Yuta Ito,$^{a}$
Jun Nishimura${}^{ab}$ and Asato Tsuchiya${}^{c}$
\vspace*{0.5cm} \\
\llap{$^a$}Department of Particle and Nuclear Physics,\\
Graduate University for Advanced Studies (SOKENDAI),\\
Tsukuba, Ibaraki 305-0801, Japan\\
\llap{$^b$}High Energy Accelerator Research Organization (KEK),\\
Tsukuba, Ibaraki 305-0801, Japan\\
\llap{$^c$}Department of Physics, Shizuoka University,\\
836 Ohya, Suruga-ku, Shizuoka 422-8529, Japan
\vspace*{0.5cm} \\
\email{yito@post.kek.jp,
jnishi@post.kek.jp,satsuch@ipc.shizuoka.ac.jp}}

\preprint{KEK-TH-1836}

\abstract{
Recent studies on the Lorentzian version of the type IIB matrix model 
show that (3+1)D expanding universe emerges dynamically
from (9+1)D space-time predicted by superstring theory. 
Here we study a bosonic matrix model obtained by 
omitting the fermionic matrices.
With the adopted simplification and the usage of a large-scale
parallel computer, we are able to perform Monte Carlo calculations
with matrix size up to $N=512$, which is twenty times larger than 
that used previously for the studies of the original model.
When the matrix size is larger than some critical value 
$N_{\rm c}\simeq 110$, we 
find that (3+1)D expanding universe
emerges dynamically with a clear large-$N$ scaling property.
Furthermore, at sufficiently late times,
we observe a power-law behavior $t^{1/2}$ 
of the spatial extent with respect to time $t$,
which is reminiscent of the expanding behavior
of the Friedmann-Robertson-Walker universe
in the radiation dominated era.
We discuss possible implications of this result
on the original model including fermionic matrices.
}

\keywords{Matrix Models, 1/N Expansion}

\usepackage{babel}

\makeatother

\newcommand {\beq} {\begin{equation}}
\newcommand {\eeq} {\end{equation}}
\newcommand {\beqa}{\begin{eqnarray}}
\newcommand {\eeqa}{\end{eqnarray}}

\begin{document}
\tableofcontents{}

\section{Introduction}

Understanding how our universe began is a fascinating
subject in theoretical physics.
This is, of course, not so easy since one has to
deal with quantum gravity near cosmic singularity, 
which appears in general relativity.
As a promising way to describe quantum gravity,
string theory has been studied for many years.
However, as far as one applies
perturbative string theory around a time dependent background,
cosmic singularity cannot be resolved 
in general \cite{Liu:2002kb,Lawrence:2002aj,Horowitz:2002mw,Berkooz:2002je}.
It is therefore necessary to study string theory
in a nonperturbative background-independent fashion.
As nonperturbative formulations
of superstring/M theory, 
matrix models that
can be obtained formally by dimensionally reducing 
10D $\mathcal{N}=1$ super
Yang-Mills theory to $d=0$ \cite{Ishibashi:1996xs},
$d=1$ \cite{Banks:1996vh} and $d=2$ \cite{Dijkgraaf:1997vv}
were proposed. 
As a closely related
direction, ref.~\cite{McFadden:2009fg,Bzowski:2012ih} proposes a conformal 
field theory, which is holographically dual to inflationary models.

The type IIB matrix model\cite{Ishibashi:1996xs} 
is one of these proposals corresponding to the $d=0$ case above.
A peculiar feature of this model
is that both space and time emerge 
dynamically from the matrix degrees of freedom. 
In this context, 
the idea of emergent gravity has been 
pursued \cite{Szabo:2006wx,Steinacker:2007dq,%
Steinacker:2008ri,Klammer:2009ku,Steinacker:2010rh,Yang:2006dk,%
Yang:2008fb,Yang:2015vna}
in gauge theories on non-commutative space 
that appear from the type IIB matrix model for a particular class 
of backgrounds \cite{Aoki:1999vr,Ambjorn:1999ts,%
Ambjorn:2000nb,Ambjorn:2000cs}. 
There are also other proposals for
the description of curved space-time 
in matrix models \cite{Hanada:2005vr,Ishiki:2015saa,deBadyn:2015sca}.

Until recently, the type IIB matrix 
model 
was studied after the ``Wick rotation'' 
\cite{Aoki:1998vn,Hotta:1998en,Ambjorn:2000bf,Ambjorn:2000dx,%
Anagnostopoulos:2001yb,Nishimura:2000ds,Nishimura:2000wf,%
Nishimura:2001sx,Kawai:2002jk,Aoyama:2006rk,Imai:2003ja,%
Imai:2003jb,Imai:2003vr,Anagnostopoulos:2013xga}
since the partition function
of the Euclidean model obtained in this way is shown to be 
finite \cite{Krauth:1998xh,Austing:2001pk}.
However, the Euclidean model is clearly not suitable
for studying the real-time dynamics
since the time coordinate is treated as purely imaginary
due to the ``Wick rotation''. 
Moreover, it is known that the Wick rotation is more subtle 
in quantum gravity
than in quantum field theories at the nonperturbative level (See,
for instance, refs.~\cite{Ambjorn:2005qt,Kawai:2011rj}). 
In fact, according to a recent study of the Euclidean model
using the Gaussian expansion method,
the emergent space-time seems to have only 
three dimensions \cite{Nishimura:2011xy}.

On the other hand, 
the Lorentzian version of the type IIB matrix model
has been studied
for the first time in ref.~\cite{Kim:2011cr}.
Unlike the Euclidean version, one has to introduce
infrared cutoffs in the temporal and spatial directions 
to make the partition function finite.
Despite this subtlety, 
the time-evolution of the ``universe'' was extracted
from the matrix configurations generated by Monte Carlo
simulation, and a large-$N$ scaling behavior
was observed with the matrix size $N \le 16$.
Quite remarkably, the SO(9) rotational symmetry of
the 9D space turned out to be 
spontaneously broken down to SO(3) at some critical time,
after which only three out of the
nine spatial directions start to expand.

In order to see what happens at later times in this model, 
one needs to increase the matrix size, 
which makes the Monte Carlo simulation more and more time-consuming.
As an alternative approach, 
one can use the classical 
approximation \cite{Kim:2011ts,Kim:2012mw,%
Stern:2014aqa,Chatzistavrakidis:2014vsa,Chaney-Lu-Stern}
to investigate possible behaviors at late 
times since each term in the action becomes large 
as the expansion of the ``universe'' proceeds.
A general prescription to construct solutions to the classical
equations of motion was given in ref.~\cite{Kim:2012mw}.
One can actually construct classical solutions corresponding to
an expanding (3+1)D universe, which naturally solve 
the cosmological constant problem \cite{Kim:2012mw}.
As a closely related progress,
it was found that
matrix configurations with intersecting fuzzy spheres in
the extra dimensions can 
accommodate
the standard model fermions \cite{Chatzistavrakidis:2011gs,Aoki:2010gv,%
Nishimura:2013rfa,Nishimura:2013moa,Aoki:2014cya,Steinacker:2014fja,%
Steinacker:2014eua}.

In fact, it is known that the classical equations of motion
of the matrix model have infinitely many solutions \cite{Kim:2012mw}.
Therefore, in order to determine which classical solution is 
actually realized at late times,
we need to study the time-evolution of the ``universe''
at least for a sufficiently long time 
by performing Monte Carlo simulation.
To that end, we previously studied
a simplified model that describes
the early time behaviors of the original model \cite{Ito:2013ywa}.
With the matrix size $N \le 64$,
we observed a clear exponentially expanding 
behavior,
which is reminiscent of the inflation.
Monte Carlo studies of the original model
with $N=24$ \cite{Ito:2013qga}
yielded results consistent with this observation.

In this paper we study a bosonic model,
which can be obtained by simply omitting the fermionic matrices.
This simplification and the usage of a large-scale parallel
computer enable us to 
perform Monte Carlo simulation
with the matrix size up to $N=512$.
Unlike the original model, the eigenvalue distribution 
of the temporal matrix turns out to have a finite extent
without introducing a cutoff
in the temporal direction.
The extent of the eigenvalue distribution is independent of 
$N$ for $N< N_{\rm c}\simeq 110$, but it increases with $N$
for $N\ge N_{\rm c}$.
We find that the properties of the model
changes drastically at the critical $N= N_{\rm c}$.
For $N < N_{\rm c}$, the dominant matrix
configurations do not allow extraction of a well-defined
time-evolution.
For $N \ge N_{\rm c}$, on the other hand, 
we can extract a meaningful time-evolution,
which shows that 
the SO(9) rotational symmetry is broken spontaneously 
down to SO(3) symmetry 
at some point in time 
similarly to
the original model.
The large-$N$ scaling behavior is clearly observed.
%
The expanding behavior of the spatial extent 
can be fitted by an exponential function 
only for a rather short period,
and after that it becomes
a power-law $t^{1/2}$ with respect to time $t$.
The latter behavior 
coincides with
the expanding behavior
of the Friedmann-Robertson-Walker universe
in the radiation dominated era.
From these results, we speculate that 
the exponential expansion of the space
in the original model suggested in the previous works
actually terminates at some point in time
and 
turns into a power law similarly to the bosonic model.
This would imply that the number of e-foldings is determined
dynamically in the Lorentzian type IIB matrix model.

The rest of this paper is organized as follows. 
In section \ref{sec:review} we briefly
review some important properties of the Lorentzian type IIB matrix model.
In section \ref{sec:bosonic-def} 
we define the bosonic model we study in this paper,
and discuss the existence of the critical $N$, at which the properties
of the model change drastically.
In sections \ref{sec:below-Nc} and \ref{sec:above-Nc}
we discuss in detail 
the properties of the model below and above $N_{\rm c}$, respectively.
Section \ref{sec:summary} is devoted
to a summary and discussions.
In Appendix \ref{sec:simulation-details}
we give the details of our simulation.
In Appendix \ref{sec:appendix} we present our results for 
the (5+1)D version of the bosonic model.

\section{Brief review of the Lorentzian type IIB matrix model}
\label{sec:review}

The action of the Lorentzian type IIB matrix 
model
is given by \cite{Ishibashi:1996xs}
\begin{eqnarray}
S & = & S_{{\rm b}}+S_{{\rm f}} \ ,
\label{eq:S_likkt}\\
S_{{\rm b}} & = & 
\frac{1}{4g^{2}}{\rm Tr}
\left(\left[A_{\mu},A_{\nu}\right]
\left[A^{\mu},A^{\nu}\right]\right) \ ,
\label{eq:Sb}\\
S_{{\rm f}} & = & 
-\frac{1}{2g^{2}}{\rm Tr}
\left(\Psi_{\alpha}\left(\mathcal{C}
\Gamma^{\mu}\right)_{\alpha\beta}
\left[A_{\mu},\Psi_{\beta}\right]\right) \ ,
\label{eq:Sf-1}
\end{eqnarray}
where the bosonic $N\times N$ matrices 
$A_{\mu}$ $\left(\mu=0,\ldots,9\right)$
and the fermionic matrices 
$\Psi_{\alpha}$ $\left(\alpha=1,\ldots,16\right)$
are both traceless and Hermitian. 
$\Gamma^{\mu}$ are 10D gamma-matrices
after the Weyl projection and $\mathcal{C}$ is the charge conjugation
matrix. The ``coupling constant'' $g$ is merely a scale parameter
in this model since it can be absorbed by rescaling $A_{\mu}$ and
$\Psi$ appropriately. 
The indices $\mu$ and $\nu$
are contracted using the Lorentzian metric 
$\eta_{\mu\nu}={\rm diag}\left(-1,1,\ldots,1\right)$.
The Euclidean version can be obtained by making the 
``Wick rotation'' $A_0 = i A_{10}$, where $A_{10}$ 
is supposed to be Hermitian.

The partition function 
for the Lorentzian version
is proposed in ref.~\cite{Kim:2011cr} as 
\begin{equation}
Z=\int dAd\Psi\, e^{iS}
\label{Z-Likkt1}
\end{equation}
with the action \eqref{eq:S_likkt}. 
The ``$i$'' in front of the action
is motivated from the fact that
the string world-sheet metric should also have 
a Lorentzian signature. 
By integrating out the fermionic matrices,
we obtain the Pfaffian 
\begin{equation}
\int d\Psi\, e^{iS_{{\rm f}}} =
{\rm Pf}\mathcal{M}\left(A\right)  \ ,
\end{equation}
which is real unlike in the Euclidean case \cite{Anagnostopoulos:2013xga}.
Note also that the bosonic action \eqref{eq:Sb}
can be written as 
\begin{eqnarray}
S_{{\rm b}}  =  
\frac{1}{4g^{2}}{\rm Tr}\left(F_{\mu\nu}F^{\mu\nu}\right)
 =  \frac{1}{4g^{2}}
\left\{ {\rm -2Tr}\left(F_{0i}\right)^{2}+
{\rm Tr}\left(F_{ij}\right)^{2}\right\} \ ,
\label{decomp-Sb}
\end{eqnarray}
where we have introduced
the Hermitian matrices $F_{\mu\nu}=i\left[A_{\mu},A_{\nu}\right]$.
Since the two terms
in the last expression
have opposite signs,
$S_{{\rm b}}$ is not positive semi-definite,
and it is not bounded from below.

In order to make the partition function \eqref{Z-Likkt1} finite,
one needs to introduce infrared cutoffs 
in both the temporal and spatial directions, 
for instance, as \cite{Kim:2011cr}
\begin{eqnarray}
\frac{1}{N}{\rm Tr}\left(A_{0}\right)^{2} 
& \leq & \kappa\frac{1}{N}{\rm Tr}
\left(A_{i}\right)^{2} \ ,\label{eq:t_cutoff}\\
\frac{1}{N}{\rm Tr}\left(A_{i}\right)^{2} 
& \leq & \Lambda^{2} \ .
\label{eq:s_cutoff}
\end{eqnarray}
It was found that the two parameters $\kappa$ and $\Lambda$ can be removed
in the large-$N$ limit in such a way that physical 
quantities scale \cite{Kim:2011cr}.
Therefore the resulting theory can be characterized by
a single scale parameter.

In the present work, it will be important to
understand the reason 
why we need to introduce the cutoff (\ref{eq:t_cutoff})
in the temporal direction.
Note first that one can use
the ${\rm SU}\left(N\right)$ symmetry of the model
to bring the temporal matrix $A_{0}$ into the diagonal form
\begin{equation}
A_{0}={\rm diag}\left(\alpha_{1},\ldots,\alpha_{N}\right)\ ,
\quad \quad
{\rm where~} \alpha_{1}<\cdots<\alpha_{N} \ .
\label{eq:diagonal gauge}
\end{equation}
By ``fixing the gauge'' in this way,
we can rewrite the partition function (\ref{Z-Likkt1}) as
\beqa
\label{gauge-fixing}
Z &=& \int  \prod_{a=1}^{N}d\alpha_{a}\,
\Delta (\alpha)^2 \, \int dA_i d\Psi\, e^{iS} \ , 
\\
\Delta (\alpha) &\equiv &
\prod_{a>b}^{N}
\left(\alpha_{a}-\alpha_{b}\right) \ ,
\label{A0diag}
\eeqa
where $\Delta(\alpha)$ is the van der Monde determinant.
The factor $\Delta (\alpha)^2$ 
in (\ref{gauge-fixing})
appears from the Fadeev-Popov procedure
for the gauge fixing, and it acts as a repulsive potential 
between the eigenvalues $\alpha_i$ of $A_0$.
Here we consider a situation in which the eigenvalues of $A_0$
are well separated from each other.
Then the action $S = S_{{\rm b}}+S_{{\rm f}}$ can be expanded as
\beqa
S_{{\rm b}} &=& -\frac{1}{2g^{2}} (\alpha _ a - \alpha _ b)^2 |(A_i)_{ab}|^2
+ \cdots \ , 
\label{Sb-leading}
\\
S_{{\rm f}} & = & 
-\frac{1}{2g^{2}} (\Psi_{\alpha})_{ba} 
(\alpha _ a - \alpha _ b)
\left(\mathcal{C}\Gamma^{\mu}\right)_{\alpha\beta}
(\Psi_{\beta})_{ab}  + \cdots \ ,
\label{Sf-leading}
\eeqa
omitting the subleading terms
for large $|\alpha _ a - \alpha _ b|$.
Integrating out $A_i$ at one loop
neglecting the zero modes corresponding to 
diagonal elements, we obtain
$\Delta(\alpha)^{-18}$.
On the other hand, integrating out $\Psi_\alpha$ at one loop
similarly, we obtain $\Delta(\alpha)^{16}$.
Thus we find that 
the potential
between $\alpha_i$
is canceled exactly at the one-loop level.
This is actually a consequence of 
supersymmetry \cite{Ishibashi:1996xs} of the model (\ref{Z-Likkt1}).
Owing to this property, the eigenvalue distribution of $A_0$ 
extends to infinity even for finite $N$ if 
the cutoff (\ref{eq:t_cutoff}) were absent.

In fact, after some manipulation and rescaling of $A_{\mu}$, 
we can rewrite the partition function (\ref{Z-Likkt1})
as \cite{Kim:2011cr} (See 
Appendix A of ref.~\cite{Ito:2013ywa} for a refined argument.)
\beqa
Z&=&\int dA\,{\rm Pf}\mathcal{M}
\left(A\right)\delta\left(
\frac{1}{N}{\rm Tr}\left(F_{\mu\nu}F^{\mu\nu}\right)\right)
\delta\left(\frac{1}{N}{\rm Tr}\left(A_{i}\right)^{2}-1\right)
\theta\left(\kappa -\frac{1}{N}{\rm Tr}\left(A_{0}\right)^{2}\right)\ ,
\nonumber
\\
 & = & \int\prod_{a=1}^{N} d \alpha_{a}
\prod_{i=1}^{d}dA_{i} \, 
\Delta^{2}(\alpha)
\,{\rm Pf}\mathcal{M}\left(A\right)
\delta\left(\frac{1}{N}{\rm Tr}
\left(F_{\mu\nu}F^{\mu\nu}\right)\right)
\nonumber
\\
&~& \mbox{~~~~~~~~~~} \times
\delta\left(\frac{1}{N}{\rm Tr}
\left(A_{i}\right)^{2}-1\right) 
\theta\left(\kappa -\frac{1}{N}{\rm Tr}\left(A_{0}\right)^{2}\right)
 \ ,
\label{Z-Likkt3}
\eeqa
where $\theta\left(x\right)$ is the Heaviside step function.
This form allows us to performing Monte Carlo simulation
of the Lorentzian model without the sign problem 
unlike the Euclidean model.\footnote{Strictly speaking,
the Pfaffian ${\rm Pf}\mathcal{M}$ 
in (\ref{Z-Likkt3})
can change its sign,
but it turned out that configurations with positive Pfaffian
dominate at large $N$.}

It turns out that
one can extract a time-evolution
from configurations generated by simulating (\ref{Z-Likkt3}).
A crucial observation is that 
the spatial matrices $A_{i}$ have
a band-diagonal structure
in the SU($N$) basis in which $A_{0}$ 
has the diagonal form (\ref{eq:diagonal gauge}).
More precisely, there exists some integer $n$ such that
the elements of spatial matrices
$\left(A_{i}\right)_{ab}$ for $\left|a-b\right|>n$ are 
much smaller than those for $\left|a-b\right|\leq n$.
Based on this observation,
we may naturally consider $n\times n$ matrices 
\begin{equation}
\left(\bar{A}_{i}\right)_{IJ}\left(t\right)
\equiv\left(A_{i}\right)_{\nu+I,\nu+J} \ ,
\label{eq:def_abar}
\end{equation}
as representing the state of the universe at time $t$, where
$I,J=1,\ldots , n$ and $\nu=0,1,\ldots , N-n$.
The time $t$ in (\ref{eq:def_abar}) is defined by 
\begin{equation}
t=\frac{1}{n}\sum_{I=1}^{n}\alpha_{\nu+I}
\label{eq:def_t}
\end{equation}
corresponding to the $n\times n$ matrices $\bar{A}_{i}$. 
For example,
we can define the extent of space at time $t$ as 
\begin{equation}
R^{2}\left(t\right)=
\left\langle \frac{1}{n}{\rm tr}\sum_{i}
\left(\bar{A}_{i}\left(t\right)\right)^{2}\right\rangle \ ,
\label{eq:def_rsq}
\end{equation}
where the symbol ${\rm tr}$ represents
a trace over the $n\times n$ block.
We also define
the ``moment of inertia tensor'' 
\begin{equation}
T_{ij}\left(t\right)
=\frac{1}{n}{\rm tr}
\left(\bar{A}_{i}\left(t\right)\bar{A}_{j}\left(t\right)\right) \ ,
\label{eq:def_tij}
\end{equation}
which is a $9\times9$ real symmetric matrix. 
The eigenvalues of $T_{ij}\left(t\right)$,
which we denote by $\lambda_{i}\left(t\right)$ with the order
\begin{equation}
\lambda_{1}\left(t\right)>\lambda_{2}
\left(t\right)>\cdots>\lambda_{9}\left(t\right) \ ,
\end{equation}
represent the spatial extent in each of 
the nine directions at time $t$.
Note that the expectation values 
$\left\langle \lambda_{i}\left(t\right)\right\rangle $
tend to be equal in the large-$N$ limit if the SO(9) symmetry is
not spontaneously broken. 
This is the case at early times of the time-evolution.
After a critical time $t_{{\rm c}}$, however,
we find that three largest eigenvalues
$\left\langle \lambda_{i}\left(t\right)\right\rangle$ 
($i=1$, $2$, $3$)
become significantly larger than the others,
which implies that
the SO(9) symmetry is spontaneously broken down to SO(3).


It would be interesting to study a long time-evolution
of the model and see how the expansion of space proceeds.
This requires very large matrices,
which makes the simulation unfeasible.
In the previous work \cite{Ito:2013ywa}, we studied 
a simplified model, in which the Pfaffian is replaced by
the one-loop contribution $\Delta(\alpha)^{16}$ mentioned above.
This replacement is expected to be valid at early times,
where the expansion of space has not proceeded much
and the leading term in (\ref{Sf-leading}) is indeed dominant.
According to the argument below (\ref{Sf-leading}),
the potential between the eigenvalues of $A_0$ is canceled 
at one loop
and hence the cutoff (\ref{eq:t_cutoff}) in the temporal direction
is needed in this simplified model as well as in the original model.
On the other hand, this simplified model can be simulated 
with much less effort than the original 
model.\footnote{In order to make one trajectory
in the Hybrid Monte Carlo algorithm,
the original model requires O($N^5$) arithmetic operations, whereas
the simplified model requires only O($N^3$) arithmetic operations.
The reason for this is that the number of iterations required for
the convergence of the conjugate gradient method
used to implement the effects of fermions grows as O($N^2$).
}
In ref.~\cite{Ito:2013ywa} 
the (5+1)D version
of the simplified model was studied with the matrix size $N \le 64$,
and the SO(5) symmetry was found to be broken 
spontaneously down to SO(3)
at some point in time similarly to the original model.
Moreover, the expanding behavior of the 3D space
turned out to be exponential,\footnote{This behavior is also 
confirmed with smaller matrix size $N\le 32$ with the aid of
a renormalization group method developed 
in the same paper \cite{Ito:2013ywa}.}
and no tendencies of slowing down were observed within
the scaling region.
Analogous behaviors were also confirmed for the (9+1)D
version of the simplified model.
In the original model, on the other hand,
the subleading term in the fermionic action (\ref{Sf-leading})
becomes important at late times as the expansion proceeds,
and hence it can affect the expanding behavior.

\section{The bosonic Lorentzian type IIB matrix model}
\label{sec:bosonic-def}

In this paper we study a bosonic model in which
the fermionic matrices are simply omitted.
The partition function is given by 
\begin{equation}
Z=\int dA\, e^{iS_{{\rm b}}} \ .
\label{eq:b_model}
\end{equation}

In section \ref{sec:review}
we reviewed an argument for the necessity
of the temporal cutoff in
the original model and 
the simplified model for early times.
In the present case of the bosonic model (\ref{eq:b_model}),
the same argument implies that
there is no need to introduce 
the temporal cutoff (\ref{eq:t_cutoff}),
and that we only need a cutoff 
(\ref{eq:s_cutoff}) in the spatial direction.
%
Corresponding to (\ref{Z-Likkt3}),
we can study the bosonic Lorentzian type IIB matrix model 
by simulating
\begin{eqnarray}
Z & = & \int dA \,
\delta\left(\frac{1}{N}{\rm Tr}
\left(F_{\mu\nu}F^{\mu\nu}\right)\right)
\delta\left(\frac{1}{N}{\rm Tr}\left(A_{i}\right)^{2}-1\right)\\
 & = & \int\prod_{a=1}^{N} d \alpha_{a}
\prod_{i=1}^{d}dA_{i}\,
\Delta^{2}(\alpha) \, 
\delta\left(\frac{1}{N}{\rm Tr}
\left(F_{\mu\nu}F^{\mu\nu}\right)\right)
\delta\left(\frac{1}{N}{\rm Tr}
\left(A_{i}\right)^{2}-1\right) \ ,
\label{eq:Z_bikkt}
\end{eqnarray}
which requires computational efforts comparable to the simplified model 
for early times reviewed in the previous section.
We have used a large-scale parallel computer
to simulate the model (\ref{eq:Z_bikkt})
with the matrix size up to $N=512$, 
which enables us to investigate a long time-evolution.
%
See Appendix \ref{sec:simulation-details}
for the details of the simulation.
%


We measure the quantity 
$\langle \frac{1}{N}{\rm Tr}\left(A_{0}\right)^{2} \rangle$,
which represents the extent of the eigenvalue distribution of $A_{0}$.
As we mentioned above, this quantity turns out to be finite
in the model (\ref{eq:Z_bikkt})
although we do not introduce a cutoff in the temporal direction
such as (\ref{eq:t_cutoff}).
In Fig.~\ref{fig:tra0} (Left) we plot
the results
against $N$.
At small $N$, it
is almost independent of $N$.
However, for $N \ge N_{{\rm c}}=112$, 
it begins to increase linearly with $N$.
Figure \ref{fig:tra0} (Right) shows the $N$ dependence
of the expectation values 
$\left\langle \lambda_{i}\left(t\right)\right\rangle $
of the nine eigenvalues of $T_{ij}\left(t\right)$ 
evaluated at $t=t_{\rm peak}$, where $R^2(t)$ becomes 
maximum.\footnote{In order to define $R^2(t)$ and 
$T_{ij}\left(t\right)$, we have
to specify the block size $n$ to be used in eq.~(\ref{eq:def_abar}).
See sections \ref{sec:below-Nc} and \ref{sec:above-Nc}
for the actual values of $n$ used to obtain the results 
in Fig.~\ref{fig:tra0} (Right).}
For small $N$, there is no significant
gap between the nine eigenvalues,
whereas for $N \ge N_{{\rm c}}$,
we observe a big gap between 
$\left\langle \lambda_{3}\left(t_{\rm peak} \right)\right\rangle $
and 
$\left\langle \lambda_{4}\left(t_{\rm peak} \right)\right\rangle $.
We will see in section \ref{sec:above-Nc}
that the SO(9) symmetry is broken down to SO(3)
after a critical time
similarly to the original Lorentzian type IIB matrix model. 


\FIGURE[t]{
\epsfig{file=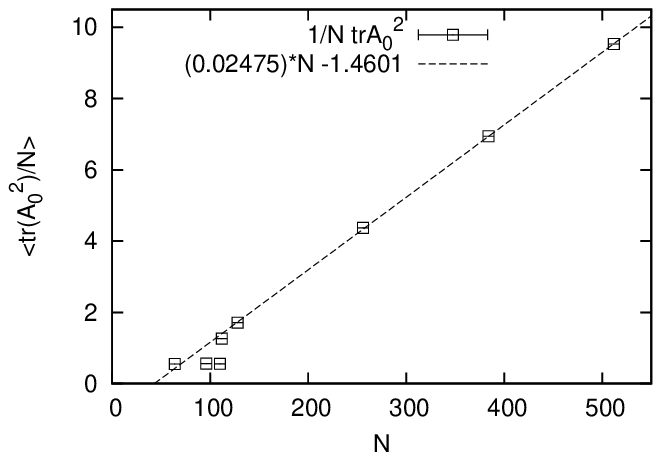,width=7.4cm}
\epsfig{file=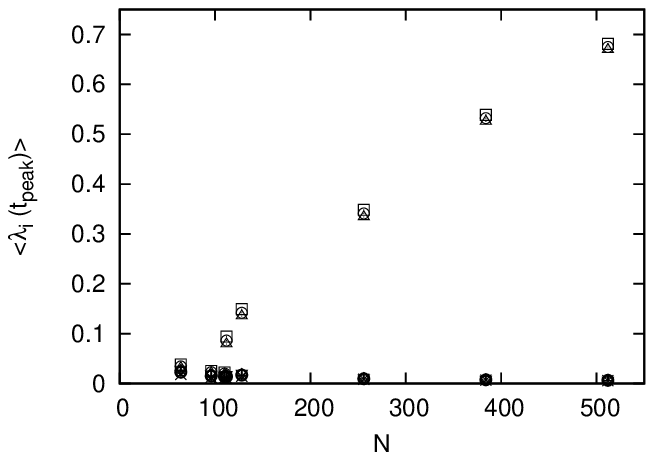,width=7.4cm}
\caption{
(Left) The extent
$\langle \frac{1}{N}{\rm Tr}\left(A_{0}\right)^{2} \rangle$
of the eigenvalue distribution of $A_0$
is plotted against $N$. 
(Right) The expectation values $\lambda_{i}\left(t\right)$
of the nine eigenvalues 
of $T_{ij}\left(t\right)$ 
at $t = t_{\rm{peak}}$ are plotted against $N$. 
For $N < N_{\rm{c}}=112$, the nine eigenvalues 
are close to each other, whereas 
for $N \geq N_{\rm{c}}$,
three out of the nine eigenvalues become much larger than the others.
}
\label{fig:tra0}
}

\section{Properties of the bosonic model for $N < N_{\rm c}$}
\label{sec:below-Nc}

In this section we discuss the properties of 
the bosonic model for $N < N_{\rm c}$. 
In order to extract the time-evolution,
we need to determine the block size $n$ to be used 
in eq.~(\ref{eq:def_abar}).
In Fig.~\ref{fig:qsq_n110} (Left)
we plot the magnitude of the off-diagonal elements of $A_{i}$
against the time separation $\alpha_{a}-\alpha_{b}$ for $N=110$.
The origin in the horizontal axis corresponds 
to the diagonal elements.
We observe a nice scaling behavior for all the matrix elements.
However, the magnitude falls off rather smoothly as one goes 
in the off-diagonal direction,
which means that the dominant matrix configurations
do not have a band-diagonal structure.

\FIGURE[t]{
\epsfig{file=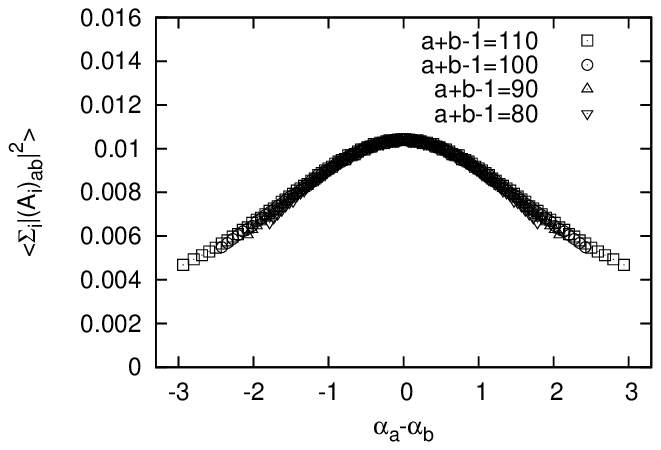,width=7.4cm}
\epsfig{file=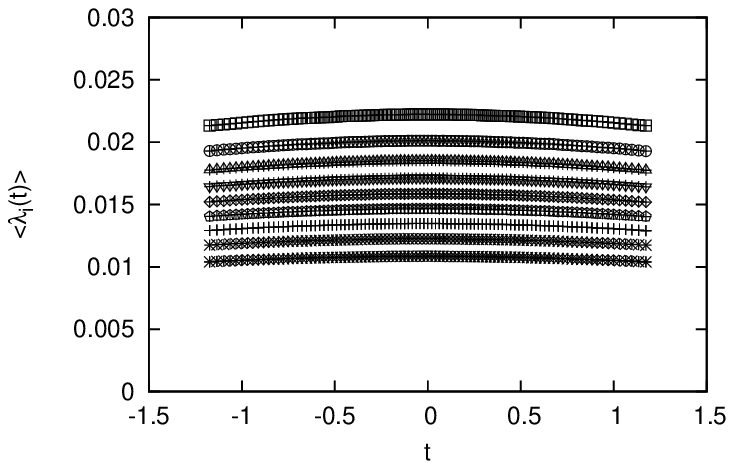,width=7.4cm}
\caption{(Left) The magnitude 
$\sum_{i}\left|\left(A_{i}\right)_{ab}\right|^{2}$ 
of the off-diagonal elements of $A_{i}$ 
is plotted against the time separation 
$\alpha_{a}-\alpha_{b}$ for $N=110$. 
(Right) The expectation values 
$\left\langle \lambda_{i}\left(t\right)\right\rangle $
of the nine eigenvalues of 
$T_{ij}\left(t\right)$ are plotted against $t$ for $N=110$.
The block size is chosen to be $n=14$.
}
\label{fig:qsq_n110}
}

In this situation, we cannot naturally define
the block matrices (\ref{eq:def_abar}) representing
the state at each time and hence 
the notion of time-evolution becomes obscure.
Let us nevertheless try to extract
the ``time-evolution'' using $n=14$ as the block size,
which is the value obtained for $N=N_{{\rm c}}=112$
in the way described in the next section.
In Fig.~\ref{fig:qsq_n110} (Right)
we plot the expectation values 
$\langle \lambda_{i}\left(t\right) \rangle$
for $N=110$.
It turns out that there is only little $t$-dependence,
and there is no clear gap between the eigenvalues for all $t$.

\FIGURE[t]{
\epsfig{file=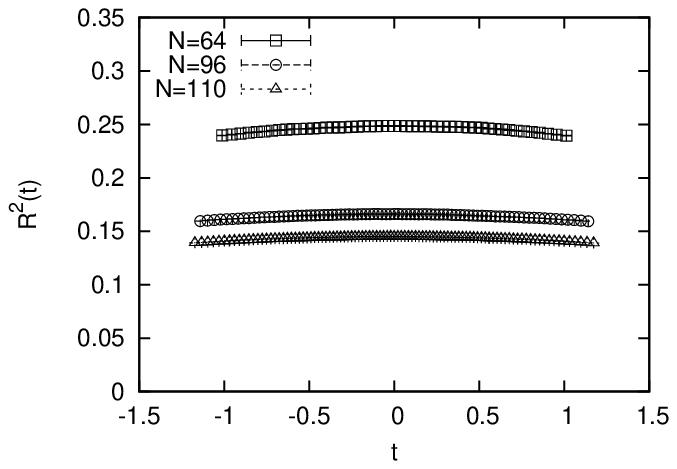,width=10cm}
\caption{The extent of space $R^{2}\left(t\right)$ is plotted against 
$t$ for $N=64$, $96$ and $110$.
The block size is chosen to be $n=14$ for all $N$.
}
\label{fig:rsq_n110}
}

The situation for smaller $N$ is similar to the $N=110$ case.
In Fig.~\ref{fig:rsq_n110}
we plot the extent of space
$R^2(t)$ as a function of $t$
for $N=64,96$ and $110$ obtained with the same block size $n=14$.
The dependence on $N$ turns out to be modest.
\section{Properties of the bosonic model for $N \ge N_{\rm c}$}
\label{sec:above-Nc}

In this section we study the properties of the bosonic model 
for $N \ge N_{\rm c}$.
In Fig.~\ref{fig:qsq_n128} (Left)
we plot the magnitude of the off-diagonal elements of $A_{i}$ 
for $N=128$.
We find that the magnitude decreases rapidly as one goes
away from diagonal elements.
Moreover, the magnitude scales only
for sufficiently large $\left|\alpha_{a}-\alpha_{b}\right|$.
From this observation, we
identify the block size $n$ as the number of points
in the region where the off-diagonal elements do not scale.
(In the present $N=128$ case, we obtain $n=20$. 
See below for more detail.) 
%
Using the block size $n$ determined in this way,
we can obtain the time-evolution.
In Fig.~\ref{fig:qsq_n128} (Right)
we plot the expectation values 
$\langle \lambda_{i}\left(t\right) \rangle$
for $N=128$. 
In contrast to the situation for $N< N_{\rm c}$, 
we observe the spontaneous symmetry breaking
from SO(9) to SO(3) at a critical time $t_{{\rm c}}$
similarly to the original Lorentzian type IIB matrix 
model.\footnote{The fact that the spatial dimensionality
after the spontaneous symmetry breaking
turned out to be the same as in the original model
is understandable from the view point of the mechanism
suggested in ref.~\cite{Kim:2011cr},
which involves only the boson part of the action.
} 

In order to study the large-$N$ scaling property,
we perform simulation for $N=256$, $384$, $512$ as well.
In Fig.~\ref{fig:qsq_n384_n512}  (Top-Left)
we zoom up the region
near the origin in Fig.~\ref{fig:qsq_n128} (Left).
From this figure,
we determine the block size for $N=128$ to be $n=20$.
Similarly, from the other figures in Fig.~\ref{fig:qsq_n384_n512},
%
we determine the block size 
for $N=256$, $N=384$ and $512$ to be 
$n=24$, $n=28$ and $32$, respectively.


\FIGURE[t]{
\epsfig{file=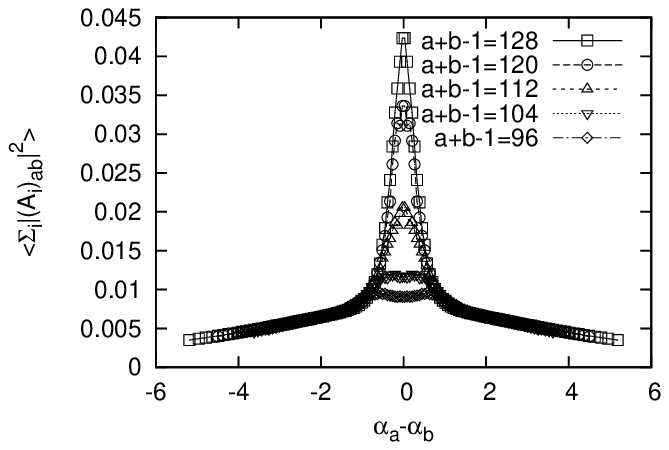,width=7.4cm}
\epsfig{file=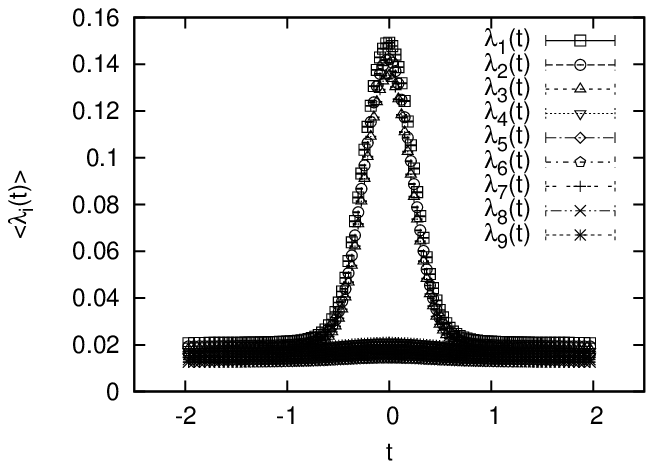,width=7.4cm}
\caption{
(Left) The magnitude $\sum_{i}\left|\left(A_{i}\right)_{ab}\right|^{2}$ 
of the off-diagonal elements of $A_{i}$ is plotted against the time 
separation $\alpha_{a}-\alpha_{b}$ for $N=128$. 
The scaling is observed only 
for sufficiently large $\left|\alpha_{a}-\alpha_{b}\right|$. 
(Right) The expectation values 
$\langle \lambda_{i}\left(t\right) \rangle$
of the nine eigenvalues of $T_{ij}\left(t\right)$ 
are plotted against $t$ for $N=128$ with the block size $n=20$. 
}
\label{fig:qsq_n128}
}


\FIGURE[t]{
\epsfig{file=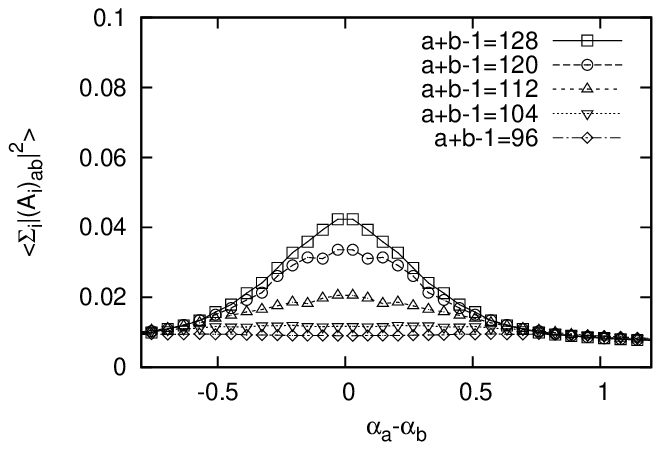,width=7.4cm}
\epsfig{file=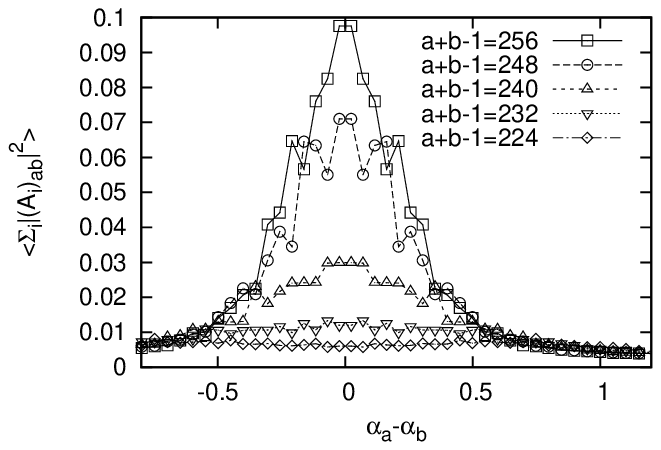,width=7.4cm}
\epsfig{file=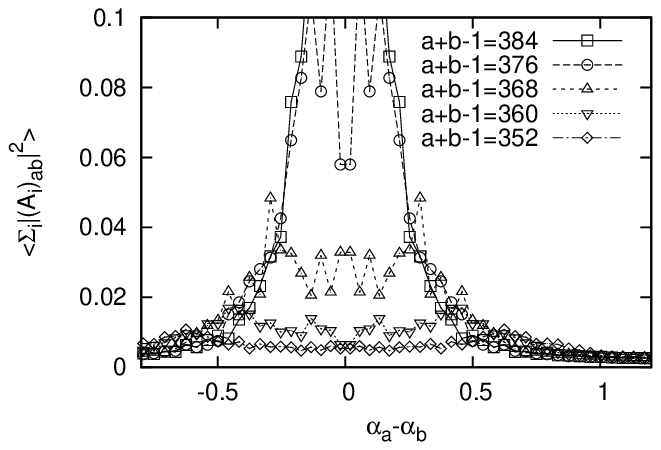,width=7.4cm}
\epsfig{file=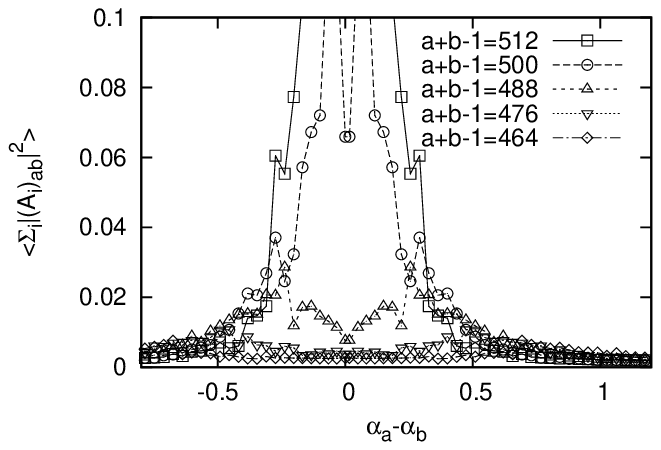,width=7.4cm}
\caption{
(Top-Left) The zoom up of the region near the origin
in Fig.~\ref{fig:qsq_n128} (Left).
We find 20 points in the region in which 
the scaling behavior is violated.
Analogous plots for $N=256$, $N=384$, $N=512$
are shown in the other panels,
where we find 24, 28, 32 points in the non-scaling region,
respectively.
}
\label{fig:qsq_n384_n512}
}

\FIGURE[t]{
\epsfig{file=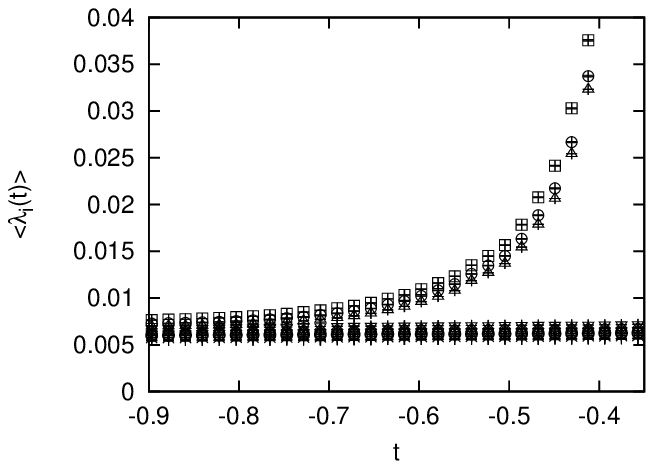,width=7.4cm}
\epsfig{file=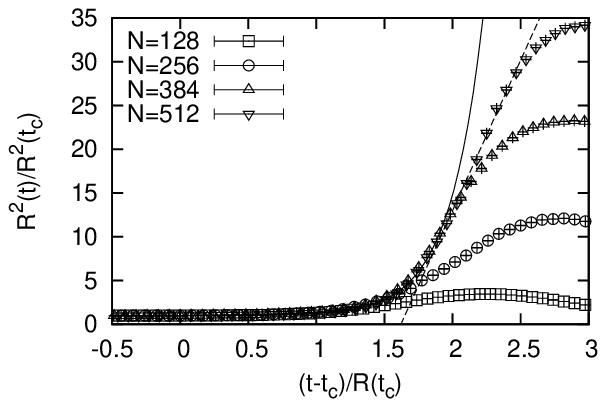,width=7.4cm}
\caption{
(Left) The expectation values 
$\langle \lambda_{i}\left(t\right) \rangle$
of the nine eigenvalues of $T_{ij}\left(t\right)$ 
are plotted against $t$ for $N=512$
with the block size $n=32$. 
(Right) The extent of space $R^{2}\left(t\right)$ 
normalized by $R^2\left(t_{\rm c}\right)$
is plotted against 
$x = \left(t-t_{\rm c}\right)/R\left(t_{\rm c}\right)$ 
for $N=128$, $256$, $384$ and $512$. 
See table \ref{tab:param_table} for the values of the block size
$n$, the critical time $t_{\rm c}$
and the extent of space $ R(t_{\rm c})$ at the critical time,
which are used to make this plot.
The solid line is a fit of the $N=512$ data to 
$R^{2}\left(t\right)/R^{2}\left(t_{\rm{c}}\right)=
a+\left(1-a\right)\exp\left(bx\right)$ 
for $ 1.0 \le x \le 1.85$,
which gives $a=0.9957(5)$ and $b=4.03(7)$. 
The dashed line is a fit of the $N=512$ data
to $R^{2}\left(t\right)/R^{2}\left(t_{\rm{c}}\right)=cx+d$ 
for $1.85 \le x \le 2.5$,
which gives $c=34.3(6)$ and $d=-55(1)$.
}
\label{fig:rsq}
}


The definition of the critical time $t_{\rm c}$ is 
ambiguous at finite $N$.
See Fig.~\ref{fig:rsq} (Left), where
we plot the expectation values 
$\langle \lambda_{i}\left(t\right) \rangle$
of the eigenvalues 
of $T_{ij}\left(t\right)$ against $t$ 
for $N=512$.
Note first that
the appearance of a gap between 
$\langle \lambda_3(t) \rangle$ and 
$\langle \lambda_4(t) \rangle$
signals the spontaneous symmetry breaking of SO(9) to SO(3).
Let us therefore
define the separation $d_j(t) = 
\langle \lambda_j(t) \rangle 
- \langle \lambda_{j+1}(t) \rangle$.
Then we find that the symmetric phase can be characterized by
$d_1(t) > d_2(t) > \cdots  > d_8(t)$,
while in the broken phase we find $d_2(t) < d_3(t)$.
Therefore we may 
define the critical time $t_{\rm c}$ by the largest value
of $t'$ such that
$d_1(t) > d_2(t) > \cdots > d_8(t)$ holds for $t \le t'$.
For instance, the critical time $t_{\rm c}$ obtained in this way
for $N=512$ from Fig.~\ref{fig:rsq} (Left)
is $t_{\rm{c}}=-0.76559(7)$.
Similarly we obtain $t_{{\rm c}}=-0.76930(7)$ for $N=384$.
Applying the same procedure to smaller $N$, 
we find that the large-$N$
scaling behavior in Fig.~\ref{fig:rsq} (Right)
becomes less clear due to finite $N$ effects.
We absorb these finite $N$ effects
by adjusting the value of $t_{{\rm c}}$ 
slightly\footnote{For $N=256$, we shift by two data points
and use $t_{\rm c}=-0.82166(6)$ instead of 
$t_{{\rm c}}=-0.76987(6)$.
Similarly, for $N=128$, we shift by four data points
and use $t_{\rm c}=-0.89472(7)$ instead of 
$t_{{\rm c}}=-0.75798(7)$.} from the one determined from the above 
procedure.
As is proposed in the original Lorentzian 
type IIB matrix model \cite{Kim:2011cr},
we use the extent of space $R(t_{\rm c})$ at the critical time
to fix the scale.
Explicit values of $R(t_{\rm c})$
are given
in table \ref{tab:param_table}
together with the block size $n$ and the critical time $t_{\rm c}$
for each $N$.

In Fig.~\ref{fig:rsq} (Right)
the extent of space $R^{2}(t)$ is plotted 
against $t$.
The large-$N$ scaling behavior is observed by
shifting the time coordinate so that the critical time
comes to the origin and by plotting dimensionful quantities
in units of $R(t_{\rm c})$.
The observed large-$N$ scaling shows that
the theory one obtains in the large-$N$ limit is 
characterized by one scale parameter $R(t_{\rm c})$
and it does not contain any dimensionless parameters.

%
%
It turns out that 
the behavior of $R^2(t)$ at $t > t_{\rm c}$
can be fitted to an exponential function
only for a finite range.
At later times, it can be fitted well by a straight line,
which corresponds to the power-law expansion
\begin{equation}
R\left(t\right)\propto t^{1/2} \ .
\label{power-law-exp}
\end{equation}
Note that this behavior is observed within the scaling region,
which implies that the suggested power law 
persists in the large-$N$ limit at least for some time region.
In Appendix \ref{sec:appendix} we present the results for
the (5+1)D version of the bosonic type IIB matrix model.
While we observe qualitatively the same behaviors,
there are also some interesting quantitative differences.

In order to understand the observed large-$N$ 
scaling further,
we investigate
how the continuum limit and the infinite volume limit
in the temporal direction are achieved 
in the large-$N$ limit.
Here we restrict ourselves to $N\ge 256$ since $N=128$
is too close to the critical value $N_{\rm c}=112$.
As the ``lattice spacing'' in the temporal direction,
we consider
the separation of data points in 
Fig.~\ref{fig:rsq} (Right) in the horizontal direction.
This quantity is actually $t$-dependent,
and it can be defined more explicitly as
$ \frac{\delta t}{R(t_{\rm c})}$,
where $\delta t$ is the difference of 
(\ref{eq:def_t}) between adjacent blocks.
In Fig.~\ref{fig:latspc} (Left) we plot
this $t$-dependent ``lattice spacing'',
choosing the horizontal axis to be the same
as in Fig.~\ref{fig:rsq} (Right).
We find clear tendency that the ``lattice spacing''
at the same point on the horizontal axis
decreases as $N$ increases.
As the ``volume'' in the temporal direction, we define
\beq
\Delta \equiv \frac{(t_{\rm peak} - t_{\rm c})}{R(t_{\rm c})} \ .
\label{def-Delta}
\eeq
Using this quantity,
we can also define an average ``lattice spacing'' 
$\varepsilon=\Delta/\nu$,
where $\nu$ is the number of data points
within the region $[t_{\rm c} , \ t_{\rm peak}]$.
The values of $\varepsilon$ and $\Delta$ obtained for each $N$
are given in table \ref{tab:param_table}.
We find that the average ``lattice spacing'' $\varepsilon$
decreases and the ``volume'' increases as $N$ becomes large.
In Fig.~\ref{fig:latspc} (Right) we plot $\varepsilon$ and
$\Delta$ against $N$ in the log scale.
The straight lines represent fits to the power-law behaviors,
although
the behaviors may be subject to a qualitative change 
at larger $N$.
In particular, it is an interesting dynamical question
whether $\Delta \rightarrow \infty$ or $\Delta \rightarrow {\rm const.}$ 
in the large-$N$ limit. 
In the former case, the expansion of space continues forever,
since the time $t_{\rm peak}$ at the peak cannot be reached
within a finite time.
In the latter case, on the other hand, the space stops expanding
in a finite time and starts to shrink.
By addressing this issue in the original model,
one can, in principle, predict the fate of our Universe.

%
%

\FIGURE[t]{
\epsfig{file=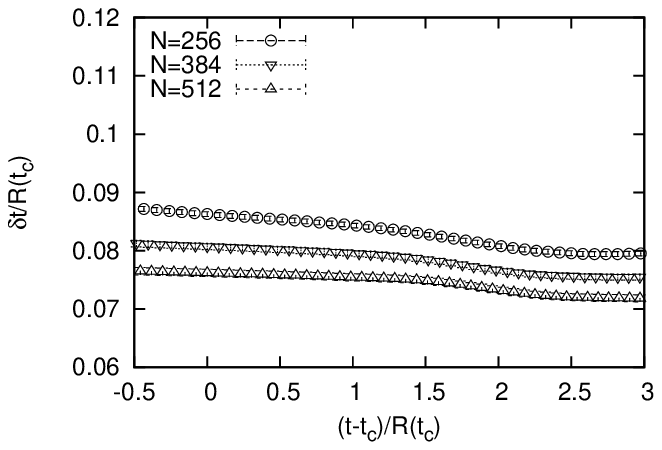,width=7.4cm}
\epsfig{file=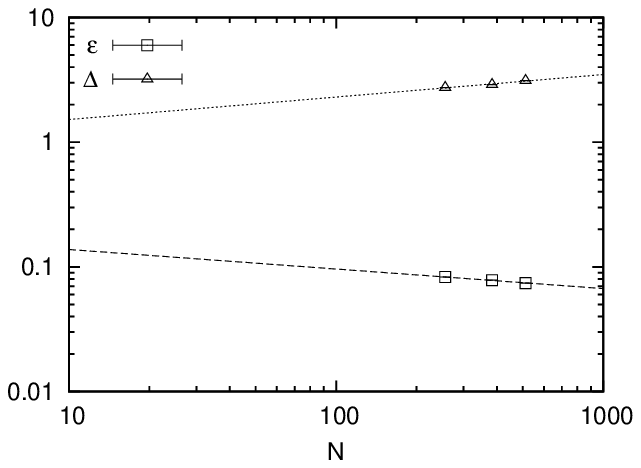,width=7.4cm}
\caption{
(Left) The ``lattice spacing''
$ \frac{\delta t}{R(t_{\rm c})}$
is plotted against $(t-t_{\rm c})/R(t_{\rm c})$.
(Right) The average ``lattice spacing'' $\varepsilon$
and the ``volume'' in the temporal direction $\Delta$
is plotted against $N$ in the log scale.
The straight lines represent fits to the power-law behaviors
$\varepsilon = a \, N^{-p}$, where $a=0.20(1)$, $p=0.16(1)$
and 
$\Delta =  b \, N^{q}$, where $b=1.0(2)$, $q=0.18(3)$.
}
\label{fig:latspc}
}

\TABLE[t]{
\centering{}%
\begin{tabular}{|c||c|c|c||c|c|}
\hline 
$N$ & $n$ & $t_{\rm c}$ & $ R(t_{\rm c})$ & 
$\varepsilon$ & $\Delta$ \tabularnewline
\hline 
\hline 
128 & 20 & -0.89472(7) & 0.39270(2) & --- & --- \tabularnewline
\hline 
256 & 24 & -0.82166(6) & 0.30045(3) & 0.08297(2)  & 2.7380(7)  \tabularnewline
\hline 
384 & 28 & -0.76930(7) & 0.26580(3) & 0.07823(2)  & 2.8943(6) \tabularnewline
\hline 
512 & 32 & -0.76559(7) & 0.24578(3) & 0.07417(2)  & 3.1150(7)  \tabularnewline
\hline 
\end{tabular}
\caption{
The block size $n$, the critical time $t_{\rm c}$
and the extent of space $ R(t_{\rm c})$ at the critical time,
which are used to make the plot
in Fig.~\ref{fig:rsq} (Right).
We also present the explicit values
of the average ``lattice spacing'' $\varepsilon$ and
the ``volume'' $\Delta$ in the temporal direction,
which are plotted in Fig.~\ref{fig:latspc} (Right).
\label{tab:param_table}
}
}



\section{Summary and discussions}
\label{sec:summary}

In this paper we have studied a bosonic model,
which can be obtained from
the Lorentzian type IIB matrix model
by omitting the fermionic matrices.
Due to the attractive potential
between the eigenvalues of $A_{0}$
arising from integrating out the spatial matrices at one loop,
the eigenvalue distribution of $A_{0}$ has a finite extent
even if one does not introduce the temporal cutoff (\ref{eq:t_cutoff}).
In the original model, this attractive potential is canceled
by the repulsive potential 
arising from integrating out the fermionic matrices at one loop
and from the van der Monde determinant.
(The simplified model for early times studied in 
ref.~\cite{Ito:2013ywa} has the same feature since the 
Pfaffian is replaced by the one-loop contribution.)
Therefore, one would naively think that supersymmetry is playing
an important role in the properties of the model that 
enable the extraction of a sensible time-evolution.

Indeed for $N <  N_{\rm c}$, we observe that the extent of 
the eigenvalue distribution of $A_{0}$ is almost independent of $N$,
and the dominant matrix configurations do not allow the
extraction of a sensible time-evolution.
However, for $N \ge  N_{\rm c}$, we find that 
the extent of the eigenvalue distribution of $A_{0}$ 
grows linearly in $N$,
and a sensible time-evolution can be extracted.
We find that the SO(9) symmetry is spontaneously 
broken down to SO(3) at some critical time similarly to the original 
model \cite{Kim:2011cr}.
The quantity such as $R(t)$ shows a clear 
large-$N$ scaling behavior.
The growth of $R(t)\propto t^{1/2}$ at late times is reminiscent
of the behavior of the Friedmann-Robertson-Walker universe
in the radiation dominated era.

Let us recall that in the simplified model for early times,
the growth of $R(t)$ was observed to be exponential \cite{Ito:2013ywa}.
In that model, only the first term in (\ref{Sf-leading}) 
was used to represent the effect of fermionic matrices.
We consider that the exponential expansion occurs 
also in the original model at early times
as is suggested by direct Monte Carlo studies
up to $N\le 24$ \cite{Ito:2013qga}.
At late times, however, the subleading term in (\ref{Sf-leading})
becomes important due to the expansion of space,
and that would affect the expanding behavior.
Note that the repulsive potential for
the eigenvalues of $A_{0}$ is obtained 
from integrating out the fermionic matrices without
the subleading term. 
Therefore one of the effects of the subleading term would be
to make the repulsive potential less effective.
Considering that the bosonic model mimics such a situation,
we speculate that the exponential expansion 
in the original model
changes into a power-law expansion at some point in time,
where the subleading term in (\ref{Sf-leading}) becomes important.
According to this scenario, the number of e-foldings is determined
dynamically in the Lorentzian type IIB matrix model.
It would be interesting to confirm the transition directly
by simulating the original model. An attempt in doing this with
a systematic approximation is in progress.

\section*{Acknowledgements}
We would like to thank K.~Anagnostopoulos,
T.~Azuma, H.~Kawai and S.-W.~Kim for discussions.
This research used computational resources 
of the K computer of the HPCI system 
provided by the AICS 
through the HPCI System Research Project (Project ID : hp130063).
The supercomputer FX10 at University of Tokyo has been
used in developing our code for parallel computing.
Computation for the (5+1)D matrix model discussed in the
Appendix \ref{sec:appendix}
was carried out mostly on PC clusters at KEK.
The work of 
Y.~I.\ is supported by Grant-in-Aid for 
JSPS
fellows.
The work of J.~N.\ and A.~T.\ is supported
by Grant-in-Aid for Scientific
Research
(No.\ 24540264, 15K05046 and 23244057)
from JSPS.

\appendix

\section{Details of Monte Carlo simulation}
\label{sec:simulation-details}

In this section we give the details on how
we perform Monte Carlo simulation 
of the bosonic model (\ref{eq:Z_bikkt}).

First the delta functions in
(\ref{eq:Z_bikkt}) are replaced by Gaussian potentials as
\begin{equation}
V_{{\rm pot}}=
\frac{1}{2}\gamma_{C}
\left(\frac{1}{N}{\rm Tr}\left(F_{\mu\nu}F^{\mu\nu}\right)\right)^{2}
+\frac{1}{2}\gamma_{L}
\left(\frac{1}{N}{\rm Tr}\left(A_{i}\right)^{2}-1\right)^{2} \ ,
\label{def-Vpot}
\end{equation}
where the coefficients $\gamma_{C}$ and $\gamma_{L}$ are
taken large enough to fix each observable to the specified value.
The values used in actual simulation are given 
in table \ref{tab:apx_table}.  

Another important issue we have to take care of is
the spontaneous breaking of the shift
symmetry $A_{0}\mapsto A_{0}+\alpha\mathbf{1}$. For instance, let us
consider calculating the expectation value $R^{2}\left(t\right)$ defined
in \eqref{eq:def_rsq}. The peak of this quantity measured
for each configuration fluctuates considerably. This reflects
the ambiguity in choosing the origin of the time coordinate, and we
should fix it before taking the ensemble average. Here we
fix it by introducing a potential
\begin{eqnarray}
V_{{\rm sym}} & = & 
\frac{1}{2}\gamma_{{\rm sym}}
\left(\frac{1}{N}\left[{\rm Tr}\left(A_{i}\right)^{2}\right]_{{\rm left}}
-\frac{1}{N}\left[{\rm Tr}
\left(A_{i}\right)^{2}\right]_{{\rm right}}\right)^{2} \ ,\\
\left[{\rm Tr}\left(A_{i}\right)^{2}\right]_{{\rm left}} & = & 
\sum_{i=1}^{d}\sum_{a+b<N+1}\left|\left(A_{i}\right)_{ab}\right|^{2} \ ,\\
\left[{\rm Tr}\left(A_{i}\right)^{2}\right]_{{\rm right}} & = & 
\sum_{i=1}^{d}\sum_{a+b>N+1}\left|\left(A_{i}\right)_{ab}\right|^{2} \ ,
\end{eqnarray}
where the values of the coefficient $\gamma_{{\rm sym}}$ used in our
simulation are given in table \ref{tab:apx_table}. 

To summarize, the model we simulate is given by
\begin{eqnarray}
Z & = & 
\int\prod_{a=1}^{N}d\alpha_{a}
\prod_{i=1}^{d}dA_{i}\, e^{-S_{{\rm eff}}} \ ,\nonumber \\
S_{{\rm eff}} & = & 
-2\log\Delta\left(\alpha\right)+V_{{\rm pot}}+V_{{\rm sym}} \ .
\label{eq:apx_z_b}
\end{eqnarray}
The simulation of the model \eqref{eq:apx_z_b}
can be performed by using the Hybrid Monte Carlo (HMC) method.
First we rewrite the model by introducing auxiliary
variables $p_{a}$ and $\left(X_{i}\right)_{ab}$$\left(a,b=1,\ldots,N\right)$
with the action
\begin{equation}
S_{{\rm HMC}}=
\frac{1}{2}\sum_{a}\left(p_{a}\right)^{2}
+\frac{1}{2}{\rm Tr}\left(X_{i}\right)^{2}
+S_{{\rm eff}}\left[\alpha,A\right] \ .
\label{eq:apx_s_hmc}
\end{equation}
Here $p_{a}$ are real variables, whereas $X_{i}$ are traceless Hermitian
matrices. We update all the variables in the model \eqref{eq:apx_s_hmc}
in the following way. 
First we regard $p_{a}$ as the conjugate momenta of
$\alpha_{a}$ and $X_{i}$ as the conjugate momenta of $A_{i}$. Then
we regard $S_{{\rm HMC}}$ in \eqref{eq:apx_s_hmc} as the Hamiltonian
$H$ and solve the classical equations of motion obtained as the 
Hamilton equations
\begin{gather}
\frac{d\alpha_{a}}{d\tau}=\frac{\partial H}{\partial p_{a}}=p_{a},
\qquad\frac{dp_{a}}{d\tau}=
-\frac{\partial H}{\partial\alpha_{a}}=
-\frac{\partial S_{{\rm eff}}}{\partial\alpha_{a}} \ ,\nonumber \\
\frac{dA_{i}}{d\tau}=\frac{\partial H}{\partial X_{i}}
=X_{i}^{*},
\qquad\frac{dX_{i}}{d\tau}=-\frac{\partial H}{\partial A_{i}}
=-\frac{\partial S_{{\rm eff}}}{\partial A_{i}} \ ,
\label{eq:apx_ham_eq}
\end{gather}
for some fictitious time $\tau$. 
This part of the algorithm is called
the Molecular Dynamics. 
In order to solve the Hamilton equations \eqref{eq:apx_ham_eq}
numerically, we discretize them using 
the so-called leap-frog discretization,
which maintains reversibility with respect to $\tau$. Starting from
the previous configuration at $\tau=0$, we obtain a new configuration
at $\tau=\tau_{{\rm f}}$ by solving \eqref{eq:apx_ham_eq} with the
step size $\Delta\tau$ so that $\tau_{{\rm f}}=N_{\tau}\cdot\Delta\tau$,
where $N_{\tau}$ is the number of steps. 
We accept the new configuration
with the probability 
$\min\left(1,\exp\left(-\Delta S_{{\rm HMC}}\right)\right)$,
where $\Delta S_{{\rm HMC}}\equiv 
S_{{\rm HMC}}\left(\tau_{{\rm f}}\right)-S_{{\rm HMC}}\left(0\right)$,
following the idea of the Metropolis algorithm to satisfy the detailed
balance. The important point here is that $S_{{\rm HMC}}$ is nothing
but the Hamiltonian $H$, which is preserved in the classical dynamics
if the equations \eqref{eq:apx_ham_eq} are solved exactly. In fact,
$\Delta S_{{\rm HMC}}$ becomes non-zero due to the discretization, 
but it is guaranteed to be a small quantity of the order 
of $\left(\Delta\tau\right)^{2}$. 
By choosing sufficiently small $\Delta\tau$, 
the acceptance rate can be kept reasonably high, which enables
the system to move around efficiently in the configuration space.
Note also 
that the auxiliary variables $p_{a}$ and $\left(X_{i}\right)_{ab}$
appear only as the Gaussian terms in \eqref{eq:apx_s_hmc}.
Therefore, we can
update them independently by using normalized Gaussian random numbers.
This procedure of refreshing the conjugate momenta should be done
each time we start a Molecular Dynamics procedure. 

To summarize, the HMC algorithm
as applied to our system can be described as follows.
\begin{enumerate}
\item Generate initial configurations of 
$p_{a}\left(0\right)$ and $X_{i}\left(0\right)$
with Gaussian distribution 
$\propto e^{-\frac{1}{2}\sum_{a}\left(p_{a}\right)^{2}}$
and $e^{-\frac{1}{2}{\rm Tr}\left(X_{i}\right)^{2}}$, respectively.
\item Evolve the fields 
$p_{a}\left(\tau\right),X_{i}\left(\tau\right),\alpha_{a}\left(\tau\right)$
and $A_{i}\left(\tau\right)$ for fictitious time $\tau_{{\rm f}}$
according to the discretized Molecular Dynamics.
\item Accept the configuration of 
$\alpha_{a}\left(\tau_{{\rm f}}\right)$
and $A_{i}\left(\tau_{{\rm f}}\right)$ 
obtained at the end of Molecular Dynamics
with the probability $\min\left(1,e^{-\Delta H}\right)$,
where $\Delta H=H\left(\tau_{{\rm f}}\right)-H\left(0\right).$
\end{enumerate}
In the HMC algorithm, there are
two parameters\footnote{These parameters can be optimized as follows. 
For fixed $\tau_{{\rm f}}$, it is optimal to choose
$\Delta\tau$ so that $\Delta\tau\times
\left(\mathrm{acceptance\: rate}\right)$
is maximized. 
Then $\tau_{{\rm f}}$ can be optimized to minimize the auto-correlation
time in units of one step in the Molecular Dynamics.
}
$\Delta\tau$ and $\tau_{{\rm f}}$.
In the present work we choose them as in table \ref{tab:apx_table}. 

\TABLE{
\centering{}%
\begin{tabular}{|c||c|c|c||c|c|c|}
\hline 
$N$ & $\gamma_{{\rm C}}/N^2$ & $\gamma_{L}/N^2$ & $\gamma_{{\rm sym}}$ & $N_{\tau}$ & $\Delta\tau$ & trajectories\tabularnewline
\hline 
\hline 
128 & 1 & 100 & 200 & 20 & 0.0015 & 2,000,000\tabularnewline
\hline 
256 & 1 & 100 & 200 & 10 & 0.0008 & 1,600,000\tabularnewline
\hline 
384 & 1 & 100 & 2,000 & 10 & 0.0004 & 1,000,000\tabularnewline
\hline 
512 & 1 & 100 & 6,000 & 10 & 0.00025 & 2,250,000\tabularnewline
\hline 
\end{tabular}
\caption{
The values of the parameters $\gamma_{{\rm C}}$,
$\gamma_{L}$ and $\gamma_{{\rm sym}}$
in (\ref{eq:apx_z_b})
used in our simulation. We also give the values
of the parameters in the HMC algorithm:
the number of steps $N_{\tau}$ in the Molecular Dynamics 
and its step size $\Delta\tau$. 
In the last column, we give the number of ``trajectories'',
which represents how many times we solve the Molecular Dynamics
after thermalization to achieve the statistics of our data.
\label{tab:apx_table}
}
}

\section{Results for the (5+1)D version of the bosonic model}
\label{sec:appendix}

In this section we present our results\footnote{Preliminary results 
shown in 
Fig.~\ref{fig:6d_rsq} (Right) 
are published in the proceedings \cite{Ito:2013qga}.}
for a bosonic model that can be obtained by omitting fermionic matrices
in the (5+1)D version of the type IIB matrix model.
The latter model is obtained formally by dimensional
reducing the 6D $\mathcal{N}=1$ super Yang-Mills theory to a point,
and it consists of six bosonic matrices 
$A_{\mu}\left(\mu=1,\ldots,6\right)$ and 
four fermionic matrices $\Psi_{\alpha}\left(\alpha=1,\ldots,4\right)$
representing four components of a 6D Weyl spinor.
The form of the bosonic part of the action is the same as 
that of the original type IIB matrix model, which is given
in \eqref{eq:Sb}.

\FIGURE{
\epsfig{file=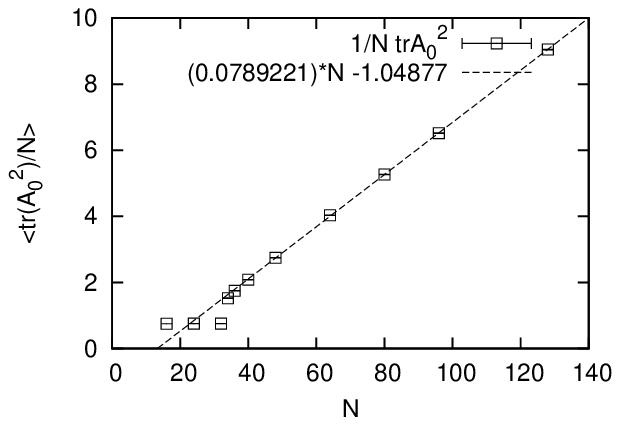,width=7.4cm}
\epsfig{file=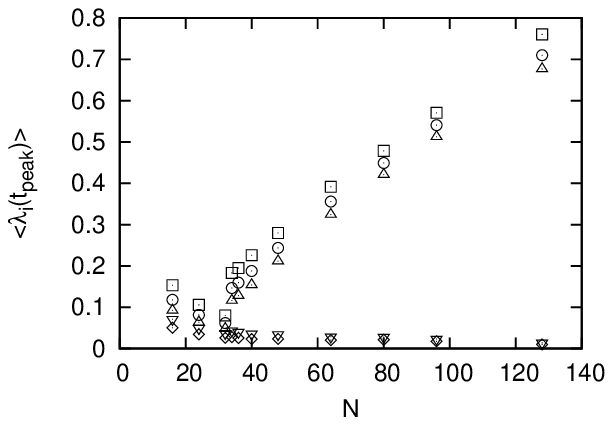,width=7.4cm}
\caption{(Left) 
The extent
$\langle \frac{1}{N}{\rm Tr}\left(A_{0}\right)^{2} \rangle$
of the eigenvalue distribution of $A_{0}$
is plotted against $N$ for the (5+1)D model.
It starts to increase at $N=N_{\rm{c}}=34$. 
(Right) The expectation values 
$\langle \lambda_{i}\left(t\right) \rangle$
of the five eigenvalues of 
$T_{ij}\left(t\right)$ at $t=t_{\rm{peak}}$ are plotted against $N$
for the (5+1)D model.}
\label{fig:6d_a0}
}

In Fig.~\ref{fig:6d_a0} (Left), we plot
the extent
$\langle \frac{1}{N}{\rm Tr}\left(A_{0}\right)^{2} \rangle$
of the eigenvalue distribution of $A_0$
against $N$.
In Fig.~\ref{fig:6d_a0} (Right), we plot
the expectation values $\lambda_{i}\left(t\right)$
of the five eigenvalues of $T_{ij}\left(t\right)$ 
at $t = t_{\rm{peak}}$ against $N$. 
While the qualitative behaviors are the same as
in the (9+1)D case shown in Fig.~\ref{fig:tra0},
we find that the critical $N_{{\rm c}}$ is smaller
and the slope of the linearly increasing
behavior of 
$\langle \frac{1}{N}{\rm Tr}\left(A_{0}\right)^{2} \rangle$
for $N\ge N_{{\rm c}}$
is larger.
We can understand this difference by considering 
the attractive potential between 
the eigenvalues of $A_{0}$
discussed below eq.~(\ref{Sb-leading}).
For general spatial dimensionality $d$,
one obtains
a factor $\Delta^{-2d}\left(\alpha\right)$ 
from integrating out the spatial matrices $A_i$ at one loop.
This factor 
acts as an attractive potential between 
the eigenvalues of $A_{0}$, 
and it is stronger for larger $d$.

\TABLE[t]{
\centering{}%
\begin{tabular}{|c||c|c|c||c|c|c|}
\hline 
$N$ & $\gamma_{{\rm C}}/N^2$ & $\gamma_{L}/N^2$ & $\gamma_{{\rm sym}}$ & $N_{\tau}$ & $\Delta\tau$ & trajectories\tabularnewline
\hline 
\hline 
64  & 1 & 100 & 1,000 & 10 & 0.0015 & 200,000\tabularnewline
\hline 
96 & 1 & 100 & 1,000  & 10 & 0.001 & 400,000\tabularnewline
\hline 
128 & 1 & 100 & 2,000 & 10 & 0.001 & 2,400,000\tabularnewline
\hline 
\end{tabular}
\caption{
The values of the parameters $\gamma_{{\rm C}}$,
$\gamma_{L}$ and $\gamma_{{\rm sym}}$
in (\ref{def-Vpot})
used in our simulation of the (5+1)D model. 
We also give the values of parameters in the HMC algorithm,
(See caption of table \ref{tab:apx_table} for explanation.)
\label{tab:apx_table_6d}
}
}

\TABLE[t]{
\centering{}%
\begin{tabular}{|c||c|c|c||c|c|}
\hline 
$N$ & $n$ & $t_{\rm c}$ & $ R(t_{\rm c})$ & 
$\varepsilon$ & $\Delta$ \tabularnewline
\hline 
\hline 
64 & 8 & -0.7248(5) & 0.1575(4) & 0.2281(6)  & 1.825(5) \tabularnewline
\hline 
96 & 10 & -0.7692(3) & 0.1276(3) & 0.2157(4)  & 2.157(4)  \tabularnewline
\hline 
128 & 12 & -0.8037(1) & 0.1070(1) & 0.2048(2)  & 2.457(2) \tabularnewline
\hline 
\end{tabular}
\caption{
The block size $n$, the critical time $t_{\rm c}$, 
the extent of space $ R(t_{\rm c})$ at the critical time,
which are used in the (5+1)D model
to make the plot in Fig.~\ref{fig:6d_rsq} (Right).
We also present the explicit values
of the average ``lattice spacing'' $\varepsilon$ and
the ``volume'' $\Delta$ in the temporal direction,
which are plotted in Fig.~\ref{fig:latspc_6d} (Right).
\label{tab:param_table_6d}
}
}

Below we discuss the properties of the (5+1)D model 
for $N\ge N_{{\rm c}}$.
(The parameters used in the simulation are listed
in table \ref{tab:apx_table_6d}.)
We have determined the block size to be $n=8$, $10$, $12$ 
for $N=64$, $96$, $128$, respectively,
from the fall-off of the off-diagonal elements of $A_{i}$ 
as is done for the (9+1)D case in section \ref{sec:above-Nc}.
In Fig.~\ref{fig:6d_rsq} (Left)
we plot the expectation values 
$\langle \lambda_{i}\left(t\right) \rangle$
of the five eigenvalues of $T_{ij}\left(t\right)$ for $N=128$,
which shows that
the SO(5) symmetry is broken spontaneously down to SO(3)
after a critical time.
From this kind of figures, 
we can determine the critical time $t_{\rm c}$ for each $N$
as described\footnote{Unlike in the (9+1)D case, 
there was no need
to adjust the value of $t_{\rm c}$ to obtain
the large-$N$ scaling behavior 
in Fig.~\ref{fig:6d_rsq} (Right).} in section \ref{sec:above-Nc}.

\FIGURE[t]{
\epsfig{file=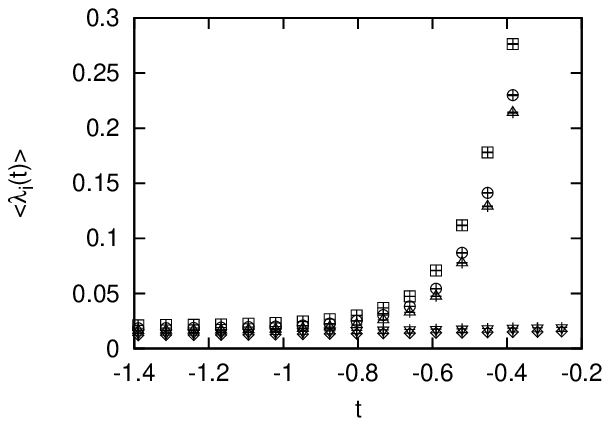,width=7.4cm}
\epsfig{file=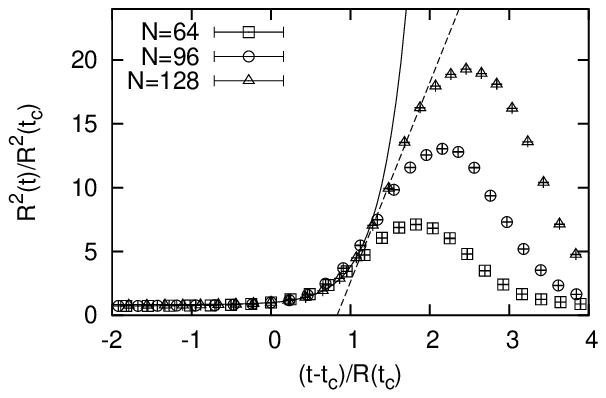,width=7.4cm}
\caption{(Left) The expectation values 
$\langle \lambda_{i}\left(t\right) \rangle$
of the five eigenvalues of $T_{ij}\left(t\right)$ 
are plotted against $t$
for the (5+1)D model with $N=128$.
The critical time is determined as 
$t_{\rm{c}}=-0.8037\left(1\right)$. 
(Right) The extent of space $R^{2}\left(t\right)$ 
is plotted against 
$x= \left(t-t_{\rm c}\right)/R\left(t_{\rm c}\right)$ 
for $N=64,96$ and $128$ 
in the (5+1)D model. 
The solid line represents a fit of the $N=128$ data to
$R^{2}\left(t\right)/R^{2}\left(t_{\rm{c}}\right)
=a+\left(1-a\right)\exp\left(bx\right)$ 
for $0.4 \le x \le 1.2$,
which gives $a=0.839(9)$ and $b=2.91(6)$. 
The dashed line represents a fit 
of the $N=128$ data to 
$R^{2}\left(t\right)/R^{2}\left(t_{\rm{c}}\right)=cx+d$ 
for $1.2 \le x \le 2.0$,
which gives $c=15.6(5)$ and $d=-13.0(8)$.}
\label{fig:6d_rsq}
}

In Fig.~\ref{fig:6d_rsq} (Right)
we show the large-$N$ scaling behavior of
the extent of space $R^{2}\left(t\right)$.
Explicit values of $R(t_{\rm c})$,
together with the block size $n$ and the critical time $t_{\rm c}$,
which are used to make this plot, are given
in table \ref{tab:param_table_6d}.
The power-law expansion (\ref{power-law-exp})
is observed at late times similarly to the (9+1)D model.

In Fig.~\ref{fig:latspc_6d} (Left)
we plot the $t$-dependent ``lattice spacing'', which shows how the 
continuum limit is achieved as $N$ increases.
The average ``lattice spacing'' $\varepsilon$ and
the ``volume'' $\Delta$ in the temporal direction
are given in table \ref{tab:param_table_6d}.
In Fig.~\ref{fig:latspc_6d} (Right) 
we plot them in the log scale.
The straight lines represent fits to the power-law behaviors.

\FIGURE[t]{
\epsfig{file=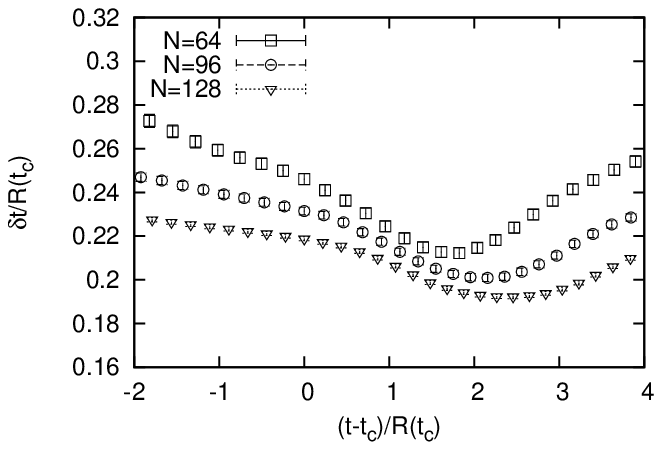,width=7.4cm}
\epsfig{file=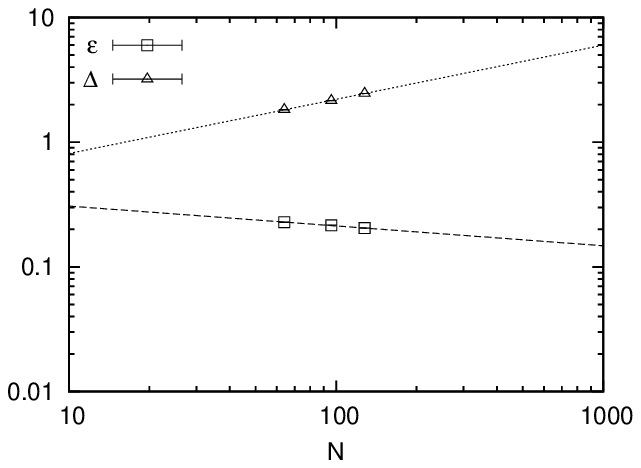,width=7.4cm}
\caption{
(Left) The ``lattice spacing''
$ \frac{\delta t}{R(t_{\rm c})}$
is plotted against $(t-t_{\rm c})/R(t_{\rm c})$ 
for the (5+1)D model.
(Right) The average ``lattice spacing'' $\varepsilon$
and the ``volume'' in the temporal direction $\Delta$
is plotted against $N$ in the log scale for the (5+1)D model.
The straight lines represent fits to the power-law behaviors
$\varepsilon = a \, N^{-p}$, where $a=0.44(2)$, $p=0.16(1)$
and 
$\Delta =  b \, N^{q}$, where $b=0.30(2)$, $q=0.43(1)$
using all the data.
}
\label{fig:latspc_6d}
}

\end{document}